\documentclass[iop]{emulateapj}
\pdfoutput=1
\usepackage{color}
\usepackage{graphicx}

\usepackage{epstopdf}

\usepackage{array}

\def\farcs{%
 \mbox{%
  \kern  0.13ex.%
  \kern -0.95ex\raisebox{.9ex}{\scriptsize$\prime\prime$}%
  \kern -0.1ex%
 }%
}%

\usepackage[hidelinks]{hyperref}
\begin{document}

\title{AGN Activity in Nucleated Galaxies as Measured by {\it Chandra}}

\author{Adi Foord\altaffilmark{1}, Elena Gallo\altaffilmark{1},  Edmund Hodges-Kluck\altaffilmark{1}, Brendan P. Miller\altaffilmark{2}, Vivienne F. Baldassare\altaffilmark{1}, Kayhan G\"{u}ltekin\altaffilmark{1}, and Oleg Gnedin\altaffilmark{1}}
\altaffiltext{1}{Department of Astronomy, University of Michigan, 1085 S. Univ., Ann Arbor, MI 48103, USA; foord@umich.edu}
\altaffiltext{2}{Department of Chemistry and Physical Sciences, The College of St. Scholastica, Duluth, MN 55811, USA}

\begin{abstract}
Motivated by theoretical expectations that Nuclear Star Clusters (NSCs) in galactic centers may provide a favorable environment for super-massive black holes to form and/or efficiently grow, we set out to measure the fraction of  nearby nucleated galaxies that also host an Active Galactic Nucleus (AGN). We targeted a distance-limited sample of 98 objects with the {\it Chandra X-ray Telescope}, down to a uniform X-ray luminosity threshold of $\sim$10$^{38}$ erg s$^{-1}$. The sample is composed of 47 late-types and 51 early-types, enabling us to further investigate the active fraction as a function of galactic morphology.  After correcting for contamination to the nuclear X-ray signal from bright X-ray binaries, we measure an active fraction $f$=11.2$\%^{+7.4}_{-4.9}$ (1$\sigma$ C.L.) across the whole sample, in agreement with previous estimates based on an heterogeneous combination of optical, X-ray and radio diagnostics, by Seth et al. (2008).  After accounting for the different stellar mass distributions in our samples, we find no statistically significant difference in the active fraction of early- vs. late-type nucleated galaxies, with $f$=10.6$\%^{+11.9}_{-4.9}$ and 10.8$\%^{+11.3}_{-6.3}$, respectively. For the early-type nucleated galaxies, we are able to carry out a controlled comparison with a parent sample of non-nucleated galaxies covering the same stellar mass range, finding again no statistically significant difference in the active fraction. Taken at face value, our findings suggest that the presence of a NSC does not facilitate nor enhance accretion-powered emission from a nuclear super-massive black hole. This is true even for late-type nucleated galaxies, home to bluer NSCs and arguably larger gas reservoirs. 
\end{abstract}



\section{Introduction} \label{sec:intro}
\par \par While all massive galaxies are thought to harbor nuclear Supermassive Black Holes (SMBHs; e.g. \citealt{Kormendy2013}), observational evidence for SMBHs becomes slim, and increasingly hard to acquire, for galaxies with stellar mass $M_{\ast}\lesssim10^{10}~M_{\odot}$. On the other hand, up to 80\% of galaxies with $M_{\ast}\lesssim10^{10}~M_{\odot}$ host a massive nuclear star cluster (NSC; \citealt{Cote2006,Georgiev2014}), with nearby early-type galaxies exhibiting a smooth transition from nuclear light deficits to nuclear light excess with decreasing galaxy mass \citep{Cote2007, Turner2012}.
NSCs are the densest known stellar concentrations in our universe, with typical half-light radii of several parsecs and masses ranging between $10^{5}M_{\odot}$ -- $10^{8}M_{\odot}$\citep{Boker2004, Walcher2005}. The resolution of the $\emph{Hubble Space Telescope}$ (HST) has allowed for detailed studies on the properties and occurrence rates of NSCs (e.g. \citealt{Carollo1997,Carollo1998,Carollo2002, Boker2002, Boker2004, Walcher2005, Walcher2006, Cote2006, Turner2012, Leigh2012, Georgiev2014,Georgiev2016}).
\par In a seminal work, \cite{Ferrarese2006} studied a large sample of nearby early-type galaxies and proposed that, as one moves down the mass function, NSCs take over from SMBHs as the main mode of compact mass aggregation in galactic nuclei. This however, does not imply that SMBHs and NSCs must be mutually exclusive, as exemplified by the nucleus of our own Milky Way \citep{Schodel2009}, as well as many others (e.g. \citealt{Seth2008,Graham2009,Gallo2010,Neumayer2012,Georgiev2014}).  In fact, theoretical work suggests that the presence of a NSC can provide a favorable environment for the formation and/or growth of massive black holes. For example, \cite{Stone2016} suggest that runaway tidal capture processes occurring in NSCs can form a $10^{2-3}~M_{\odot}$ SMBH seed which then grows via tidal disruption events to larger sizes (see \citealt{MillerD2012} and \citealt{Gnedin2014} for earlier arguments on these same lines).  An interesting case, where the above formation scenario does not seem to apply is Henize 2-10, where a NSC seems to be forming independently around a massive central black hole with stellar mass $M_{\ast}\lesssim10^{7}~M_{\odot}$ \citep{Reines2011, Nguyen2014, Arca2015}. Co-evolution between the NSC and SMBH in systems such as Henize 2-10 could occur during galaxy mergers, as modeled by \cite{Hopkins2010a, Hopkins2010b}.

In addition to facilitating the formation of a SMBH, recent star formation in NSCs could enhance the SMBH feeding-rates, and thus the expected active fraction\footnote{Here defined as the measured fraction of galaxies hosting an Active Galactic Nucleus (AGN) down to some arbitrarily defined accretion luminosity.} (\citealt{Balda2014}; see \citealt{Naiman2015} for a discussion on NSCs enhancing SMBH feeding-rates during galaxy mergers).  This effect would be even more pronounced in the nuclei of late-types, home to larger gas reservoirs.  From a different angle, \cite{Antonini2015} investigate the occurrence of NSCs and SMBHs (regardless of their accretion-powered activity level) using semi-analytical galaxy formation models (see also \citealt{Antonini2012,Antonini2013}). Among galaxies of all types, they find that the local fraction of systems containing both a NSC and SMBH ranges between a few up to 25\% for host galaxy stellar masses between $10^{9}M_{\odot}$ -- $10^{11}M_{\odot}$ (see their figure 7), with the early-types having a somewhat higher fraction.  These occupation values may represent upper limits to the measurable active fraction, allowing a comparison with observational results.
\par The most complete observational study to date on the coexistence of NSCs and actively accreting SMBHs is presented in \cite{Seth2008}, where they inspect a sample of 176 nucleated galaxies spanning all galaxy types. Of these, 75 had suitable optical spectroscopic data with either the Sloan Digital Sky Survey (SDSS) or the Palomar Survey \citep{Fili1985}; a fraction of the remaining objects had heterogeneous X-ray or radio coverage. \cite{Seth2008} find that, of the nucleated galaxies with optical spectroscopic data, 10\% are classified as AGN based on the standard optical emission line ratio BPT (``Baldwin, Phillips, \& Terlevich")  diagram \citep{Baldwin1981,Kewley2006}, with an additional 15\% falling in the AGN-star forming composite region. From this, they conclude the 
{\it active fraction of nucleated galaxies is consistent with the galaxy population as a whole}. 
\par In this paper, we search for the presence of low-luminosity AGN (i.e.\ weakly accreting SMBHs) in nucleated galaxies by means of a uniform, luminosity limited X-ray survey. Whereas optical emission line ratio diagnostics are arguably best suited at probing Eddington-ratios in excess of 10$^{-3}$, high spatial resolution X-ray imaging within $\sim$ 30 Mpc affords AGN detectability down to two orders of magnitude deeper Eddington ratios, granted a careful assessment of the X-ray binary contamination (see, e.g. \citealt{Gallo2010}).  In what follows, we report on a comparative analysis of 2 samples of nearby nucleated galaxies, both targeted with the {\it Chandra X-ray Telescope} down to a limiting luminosity of 10$^{38}$ erg sec$^{-1}$. The first sample consists of 51 nucleated Virgo early-types (for which results were first reported in \citealt{Gallo2008, Gallo2010}, \citealt{Miller2012b}); the second sample consists of 47 nucleated late-types, drawn from \cite{Seth2008} as described below. 
Our aims are two-fold: (i) to measure the overall active fraction down to a uniform limiting luminosity, and compare it with the active fraction as measured for a parent sample of non-nucleated systems over a comparable stellar mass range; (ii) to compare the active fraction as measured for nucleated early- vs. late-types. The former goal addresses whether the presence of a NSCs enhances the likelihood of there being an actively accreting SMBH; the latter will help us determine the effect  of its several-pc scale environment, if any, on SMBH feeding.


\section{Sample Selection and Data Analysis} \label{sec:data}
Out of 176 galaxies that compose the initial \cite{Seth2008} sample of nucleated galaxies, 51 are Virgo early-type; those were observed by {\it Chandra} as part of a large Cycle 8 program (AMUSE-Virgo; Proposal ID: 08900784, PI: Treu); the results were presented in \cite{Gallo2008, Gallo2010}. The remaining objects are late-type galaxies spanning distances between 4 and 40 Mpc. Out of those, we culled a sub-sample of objects (i) within $\simeq$20 Mpc; (ii) face-on, and (iii) located more than $\pm$30$^{\circ}$ off the galactic plane, so as to minimize local and Galactic absorption.  Visual inspection of optical data was used to determine the inclination of each galaxy.  This yields 47 late-type galaxies.  
The distance cut was chosen to ensure that the planned X-ray luminosity detection threshold would match that of the Virgo early-type sample that were previously targeted by {\it Chandra}. In the following, we report on the new results on the late-type galaxies. 

Out of 47 objects, 18 had sufficiently deep archival data. We targeted the remaining 29 galaxies in Cycle 16 (Proposal ID:16620343, PI: Gallo, hereafter C16 sample).  Each C16 galaxy was placed on the back-illuminated S3 chip of the Advanced CCD Imaging Spectrometer (ACIS) detector, with exposure times ranging from 1--5 ks, chosen in order to achieve a uniform 3$\sigma$ point-like source detection threshold of 10$^{38}$ erg s$^{-1}$. Typical exposure times for the archival subsample are longer than those of the C16 data, with the highest exposure being 90 ks for NGC 5879. For archival sources with  multiple visits, we prioritize the observation with the target being closer to the aim-point.  In the case of multiple observations with comparable aim-points, we prioritize the shortest exposure, again to match the sensitivity of the C16 data. \\

\par We follow a similar data reduction for the late-type galaxies as described in previous X-ray studies of the AMUSE-Virgo/Field surveys \citep{Gallo2008, Gallo2010, Miller2012a, Miller2012b, Plotkin2014, Miller2015}, using the Chandra Interactive Analysis of Observations ({\tt CIAO}) v4.8. We first correct for astrometry, preferentially cross-matching the {\it Chandra}-detected point-like sources with the Sloan Digital Sky Survey Data Release 9 (SDSS DR9) catalog, and using the U.S. Naval Observatory's USNO-B1 catalog for all other galaxies outside of the SDSS footprint. The {\it Chandra} sources used for cross-matching are detected by running {\tt wavdetect} on the reprocessed level 2 event files. We require each matched pair to be less than 2 arsec from one another, and a minimum of 3 matches for an image to be astrometrically corrected. Seventeen of the late-type galaxies are astrometrically corrected, all with shifts less than 0\farcs5. Of those, 82$\%$ use the SDSS DR9 catalog and 18$\%$ use the USNO-B1 catalog.

We then correct for background flaring by removing intervals where the background rate was found to be 3$\sigma$ above the mean level. 
We rerun {\tt wavdetect} on the filtered 0.5--7 keV data to generate a list of X-ray point sources using wavelets of scales 1, 1.5, and 2.0 pixels using a 1.5 keV exposure map, and set the detection threshold significance to 10$^{-6}$ (corresponding to one false detection over the entire S3 chip). Source position errors are calculated as the quadratic sum of the positional uncertainty for the X-ray source (discussed in \citealt{Garmire2000}), the uncertainty in the optical astrometry ($<$ 0\farcs1 for both SDSS and USNO-B1), and the uncertainty in the X-ray bore-sight correction. 

\par Next, we screen each image for diffuse X-ray emission following the methodology presented in \cite{Plotkin2014} (albeit contamination from hot, free-free emitting gas is expected to be small in our sample of nucleated late-types).  
We first generate the (source-free) galaxies' X-ray surface brightness profiles in the 0.5--7 keV band. This is done with annuli centered on the optical center of the galaxy and extending to a radius of R$_{25}$, defined as the radius where the surface brightness drops to 25 mag arcsec$^{-2}$ (R$_{25}$ values are tabulated in \citealt{Liu2011} and \citealt{Georgiev2014}). The background surface brightness, $c_{bk}$, is determined by taking the median value from 4 circular regions outside each galaxy's R$_{25}$ value. A galaxy is classified as having diffuse emission if the inner (i.e., within R$_{25}$) surface brightness profile exceeds 3$\times c_{bk}$. Two archival galaxies, NGC 4030 and NGC 5879, meet this criterion (not surprisingly, both belong to the high-end tail of the sample's mass distribution, with stellar masses log(M/M$_{\odot}$) of 10.9 and 10.0, respectively, and also have longer exposures times, of 14 and 90 ks, respectively). For these two objects, we restrict our X-ray point source search to energies between 2-7 keV, where diffuse gas typically contributes less than 5\% of the total X-ray emission, and rerun {\tt wavdetect} using a 4.5 keV exposure map. 

\par For each galaxy we search for an X-ray point source within 2$\arcsec$ of the nominal, SDSS-listed optical center (2$\arcsec$ corresponds to 95\% of the encircled energy radius at 1.5 keV for ACIS).  Counts are extracted from a 2$\arcsec$ radius circular region centered on the optical center position/X-ray source center (in the case of non detection/detection, respectively), using a source-free annulus with inner radius of 20$\arcsec$ and outer radius of 30$\arcsec$ for background extraction.  Source net count rate and flux values (or upper limits) are determined using the {\tt CIAO} script $\tt srcflux$ assuming an absorbed powerlaw spectrum (``xspowerlaw.pow1" and``xsphabs.abs1") with photon index $\Gamma$ = 1.8, and using the tool {\tt colden}\footnote{{\tt Colden} uses the National Radio Astronomy Observatory (NRAO) dataset supplied by \cite{Dickey1990}.} to evaluate the neutral hydrogen column density towards each object.
Out of 47 late-type galaxies, 5 are associated with a statistically significant, point-like, nuclear X-ray source. 
The nuclear X-ray luminosity values and upper limits are shown in Table~\ref{tab:1}, while more detailed information about the 5 detections are given in Table~\ref{tab:2}.

\section{Origin of the nuclear X-ray Emission} \label{sec:analysis}

\subsection{X-ray binary luminosity assessment}
We will now focus on the 5 late-type galaxies with nuclear X-ray detections in order to assess the physical nature of the emission. 
Based on the $\log N$--$\log S$ relation presented in \cite{Rosati2002}, the probability that any of the detected X-ray sources can be attributed to a background object within the {\it Chandra} Point Spread Function (PSF) varies between 10$^{-3}$--10$^{-4}$, and is thus deemed negligible.
The X-ray luminosities of the detected nuclei range between 0.2 -- 1 $\times$ 10$^{39}$ erg sec$^{-1}$. While nuclear X-ray luminosity in excess of 10$^{40}$ erg  sec$^{-1}$ can be comfortably attributed to AGN emission (see, e.g., \citealt{Lehmer2010}), here, the inferred values warrant a careful assessment of the contamination from bright X-Ray Binaries (XRBs) to the {\it Chandra} PSF (see \citealt{Gallo2008, Gallo2010, Miller2012b} for further discussion).  Our assessment of the origin of the detected nuclear X-ray emission will depend on the number of central ($<$2\arcsec) X-ray binary sources expected from the entire late-type sample.  To this end, we first estimate the {\it total}, expected X-ray luminosity due to XRBs for each of our 47 late-type galaxies. 
More specifically, the X-ray Luminosity Function (XLF) of Low-Mass X-ray Binaries (LMXBs) is known to be set by the galaxy cumulative star formation history, and thus scales with the total stellar mass of the galaxy, $M_{\ast}$ \citep{Ghosh2001,Grimm2002, Gilfanov2004, Kim2004, Humphrey2008}, whereas the High-Mass X-ray Binary (HMXB) XLF traces recent star formation within the galaxy \citep{Sunyaev1978, Grimm2003, Lehmer2010, Mineo2012}, and its normalization depends on the galaxy's Star Formation Rate (SFR). 
\par We start by adopting the analytic prescription by \cite{Lehmer2010}, who built on the above mentioned, previous investigations to carry out a systematic assessment of the different possible contributions to the total X-ray luminosity of nearby galaxies, including LMXBs and HMXBs, as well as diffuse emission from hot gas.  For a given $M_{\ast}$ and SFR, the total, 2--10 keV luminosity from XRBs ($L^\mathrm{gal}_\mathrm{XRB}$) can be expressed as: 

\begin{equation} 
\label{eq:1}
L^{gal}_{XRB} =  \alpha M_{*} + \beta SFR,
\label{eq:lehmer}
\end{equation}
with best fitting parameters $\alpha$$\,=\,$$(9.05 \pm 0.37) \times 10^{28}\,\mathrm{erg\,s^{-1}}\,M^{-1}_{\odot}$ and $\beta$$\,=\,$$(1.62 \pm 0.22) \times 10^{39}\, \mathrm{erg\,s^{-1}(}\,M_{\odot}\,\mathrm{yr^{-1})}^{-1}$.

\par  Stellar masses were taken from \cite{Seth2008} and were derived using galaxy colors and the $M$/$L$ relations presented in \cite{Bell2003} (errors are on the order of 0.15 dex). The formalism presented in \cite{Bell2005} was adopted to estimate SFRs, where the 12$\mu$m ($f_{12}$), 25$\mu$m ($f_{25}$), 60$\mu$m ($f_{60}$), and 100$\mu$m ($f_{100}$) flux density values from the IRAS Faint Source Catalog \citep{Moshir1990} were employed to estimate the total infrared luminosity, $L_{IR}$, for each galaxy, following \cite{Perault1987} (see also \citealt{Sanders1996}).
We use Wide-field Infrared Survey Explorer (WISE) data and IR color diagnostics (after \citealt{Jarrett2011},  \citealt{Stern2005}) to test for a possible contribution from the central SMBH to the IR luminosity for our 5 detections. We find that none of our sources fall within the AGN parameter space empirically identified by \cite{Jarrett2011}, implying that the above SFRs are likely unaffected by the presence of a central SMBH (this is not surprising, considering the extremely low Eddington ratios we are operating at). 

Five late-type galaxies (NGC 3423, NGC 3501, NGC 4183, NGC 5023, and UGC 12732) have no IRAS coverage, and we estimate their SFRs via 70 $\mu$m measurements from the $Spitzer$ Multiband Imaging Photometer (MIPS; for conversion measurements between MIPS and IRAS filters see \citealt{Terrazas2016}); 1.4 GHz measurements from the Very Large Array (VLA; for conversion measurements between 1.4 GHz and total IR emission see \citealt{Bell2003}); and H$\alpha$ measurements from the H$\alpha$ Galaxy Survey (see \citealt{James2004}), preferentially in that order.

Finally, multiple distances are available in the literature for each galaxy, including \cite{Seth2008}, \cite{Georgiev2014}, \cite{Boker2002}, and \cite{Carollo1997,Carollo1998,Carollo2002}, with the study by \cite{Seth2008} using Virgo infall-corrected velocities from Hyperleda and redshift-independent indicators listed in NED\footnote{https://ned.ipac.caltech.edu/}.  We compared the distances published in \cite{Seth2008} to those published in \cite{Georgiev2014} and the earlier publications, and find that the values agree with one-another with an expected scatter and no systematic shifts. Thus, for the purpose of drawing a fair comparison between the late and early-type sample, we adopt the distances published in \cite{Seth2008} (we verified, however, that our results are qualitatively insensitive to the distance source choice). Distances and stellar masses for each host galaxy can be found in Table \ref{tab:1}.

\subsection{Nuclear X-ray binary contamination}
In order to estimate possible contamination to the \emph{Chandra} PSF, we (i) estimate the {\it fractional} contribution to $L^\mathrm{gal}_\mathrm{XRB}$ (Eq.~\ref{eq:lehmer}) to the actual, detected nuclear X-ray signal, and (ii) estimate the total number of expected central ($<$ 2$\arcsec$) XRBs in our late-type sample. Values of the fractional contribution of $L^\mathrm{gal}_\mathrm{XRB}$ for each galaxy allows us to derive the proper normalization constants when calculating the expected number of central XRBs. Regarding point (i), we first convert the estimated $L^\mathrm{gal}_\mathrm{XRB}$ into (0.5-7) keV luminosities using webPIMMS\footnote{heasarc.gsfc.nasa.gov/cgi-bin/Tools/w3pimms/w3pimms.pl} for a power law spectrum with $\Gamma$ = 1.8. In order to estimate the mean XRB luminosity in the nucleus, we follow the methodology presented in \cite{Alfvin2016} and assume that the total $L^\mathrm{gal}_\mathrm{XRB}$ is distributed according to the optical light profile.

\begin{figure}[t]
\center
\includegraphics[width=8.4 cm]{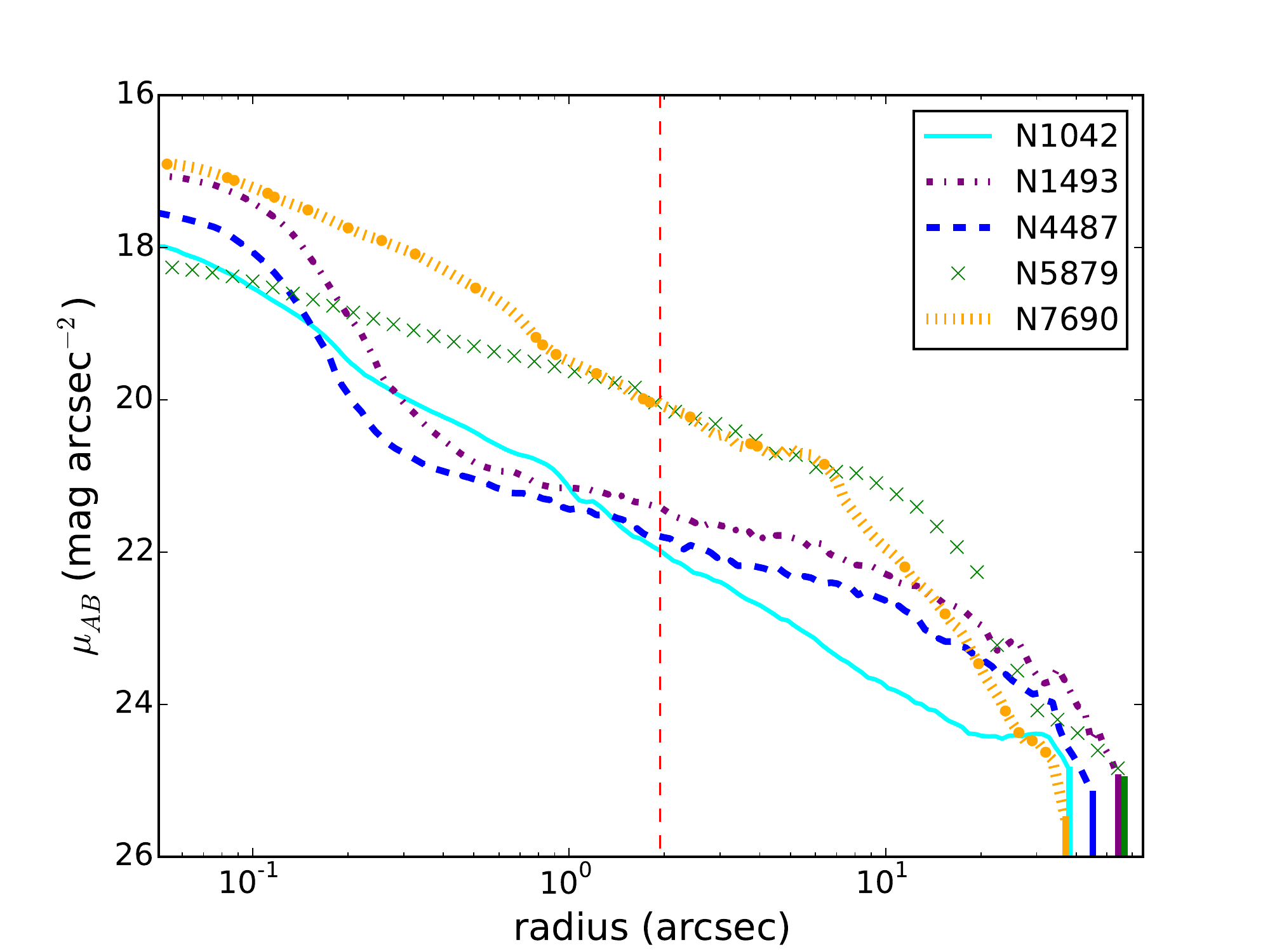}
\caption{Radial surface brightness profiles of the 5 late-type galaxies with detected nuclear X-ray emission, projected along the semi-major axis. Profiles were extracted from HLA images in the F606W (NGC1042, NGC 4487, NGC 5879, and NGC 7690) and F814W filter (NGC 1493). The red vertical dashed line marks 2$\arcsec$. For each galaxy, the solid vertical line at the end of the profile marks the extent of R$_{25}$. Error bars are $<$ 0.1 mag. }
\label{fig:latetype}
\end{figure}

%
Most of our late-type galaxies with a detected X-ray nucleus have Wide-Field Planetary Camera 2 (WFPC2) observations in the Hubble Legacy Archive (HLA)\footnote{http://hla.stsci.edu/}, where we use observations taken with the well-calibrated F606W /F814W filters. For those with no WFPC2 observations (NGC 3501, NGC 4144, NGC 4183, NGC 4206, and IC 5052) we use ACS F606W/F814W imaging. With the Image Reduction and Analysis Facility ({\tt IRAF}) software, we mask out bright foreground stars in the field of view and use {\tt ds9} to extract the radial profiles of each galaxy's surface brightness.  We integrate the light profiles out to R$_{25}$ to calculate the total luminosity of the galaxy and estimate the fraction that is contained within the central 2$\arcsec$. The radial surface brightness profiles for the 5 late-type detections, analyzed in {\tt IRAF} via {\tt ELLIPSE}, are shown in Figure~\ref{fig:latetype}. To assess for errors introduced by using different filters, we compare the fractional luminosity values (within 2$\arcsec$) between the F606W and F814W radial profiles for a handful of nearby galaxies with observations available in both bands and find the difference consistently lower than 0.4\%.
The X-ray luminosity due to XRBs within the central 2$\arcsec$ is obtained by multiplying the 0.5-7 keV total X-ray binary luminosity, $L^\mathrm{gal}_\mathrm{XRB}$, by the galaxy's respective fractional value as estimated above.  The fractions range from 0.5 -- 9.8\%, with nuclear XRB luminosities,  $L^n_\mathrm{XRB}$  between 3$\times 10^{35}$ up to $\sim$ 3$\times10^{38}$ erg s$^{-1}$.
\par Regarding point (ii), we then estimate the total number of expected central ($<$ 2$\arcsec$) XRBs in our late-type sample. For each galaxy we calculate N($>$L), where for our non-detections N is defined as the number of expected central XRBs with luminosities greater than or equal to the sensitivity threshold ($L$ $\ge$ 10$^{38}$ ergs s$^{-1}$) and for our 5 detections, N is defined as the number of expected central XRBs with luminosities greater than or equal to the detected central X-ray luminosity ($L$ $\ge$ $L_{X}$).  We assume the nuclear XRB luminosity is due to a distribution of both LMXBs and HMXBs that follow the XLFs presented in \cite{Gilfanov2004} and \cite{Grimm2003}: $dN/dL$ = K$_{1}$$L^{-\alpha1}$ + K$_{2}$$L^{-\alpha2}$, where for our given luminosity range $\alpha$1 = 1.8 and $\alpha$2 = 1.61. The normalization constants K$_{1}$ and K$_{2}$ are derived from the expected fractional XRB luminosity (see Eq.~\ref{eq:lehmer}). We build a distribution of expected nuclear XRBs from the entire late-type group by sampling through each galaxy 10$^{6}$ times and randomly pulling a value, $x$, from a Poisson distribution centered on N. If $x \ge 1$, we count $x$ XRBs from the central 2$\arcsec$ for a given galaxy. Furthermore, we assign each XRB an X-ray luminosity via inverse transform sampling from the galaxy's cumulative distribution. This process allows us to measure the most likely number of central XRBs from our sample as well as calculating the probability that, at a given luminosity, each of our central detections are SMBHs ($P_\mathrm{SMBH}$).  Following this methodology, the most likely number of total, central, XRBs in the late-type sample is projected to be $\sim$1.4, with probabilities $P_\mathrm{SMBH}$ for our five detections ranging from 94.4 to 98.1\%. \\
\begin{figure}[t]
\center
\includegraphics[width=8.4cm]{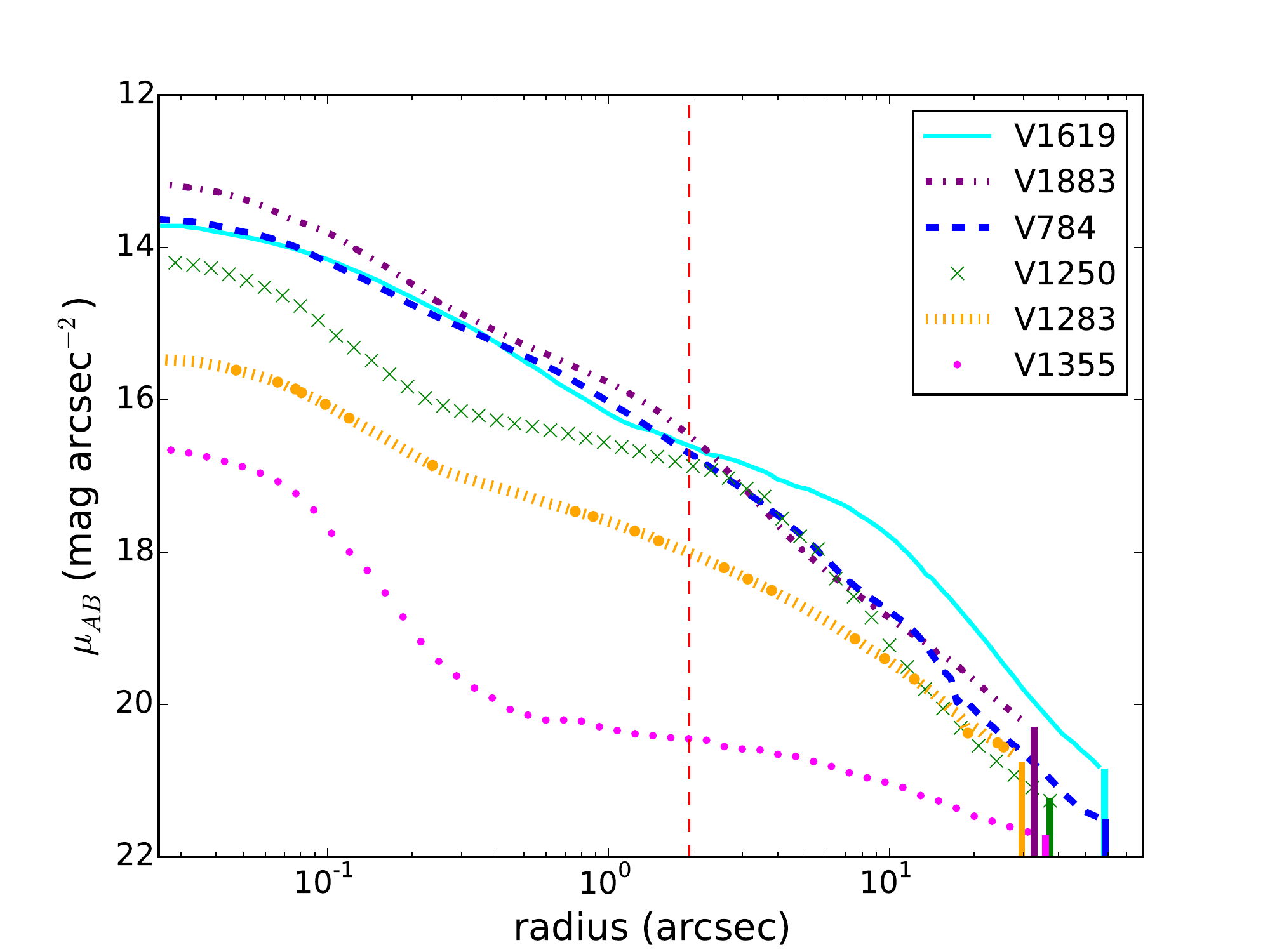}
\caption{Same as Figure~\ref{fig:latetype}, for the 6 early-type Virgo galaxies with detected nuclear X-ray emission. Profiles were extracted from HST ACS F850LP images. 
}
\label{fig:earlytype}
\end{figure}

To allow for a proper comparison between the late-type sample considered above and the companion early-type sample targeted by {\it Chandra} as part of the AMUSE-Virgo program, we carry out the same steps as described in the previous sub-section for the early-type galaxies (specifically, we wish to carry out a direct comparison for the XRB contamination, which was addressed somewhat differently by \citealt{Gallo2010}\footnote{\cite{Gallo2010} estimated the probability of XRB contamination for the nucleated Virgo galaxies using the XLF of globular clusters, as presented by \cite{Sivakoff2007}. This was motivated by the stellar number density of NSCs being arguably closer to that of globular clusters than the field; however, this approach requires knowledge of the clusters' $g-z$ colors, and hence dual band, $g$ and $z$-equivalent HST imaging data, which are not available for our sample of early types.}).  All of the 51 nucleated early-types have dual-band (F475W \& F850LP) HST Advanced Camera for Surveys (ACS) imaging data, as part of the ACS Virgo Cluster Survey \citep{Cote2006}; 6 out of 51 were found to have a nuclear X-ray source (\citealt{Gallo2010}; see Table~\ref{tab:earlytypes}). 

The radial surface brightness profiles of the six detections are shown in Fig~\ref{fig:earlytype}. 
We assess for errors introduced by using different instruments/filters between the late-type and early-type radial profiles by comparing the fractional luminosity values (within 2$\arcsec$) between the WFPC2 and ACS observations.  We find that the fractional luminosity in the F814W band is consistently 1.36  times greater than the fractional luminosity in the F850LP band, and thus multiply all F850LP fractions by this value. This fractional value is then multiplied by the estimated $L^\mathrm{gal}_\mathrm{XRB}$ in order to estimate the expected XRB luminosity from the central 2$\arcsec$. We note that SFRs for this sample\,---\,and hence the HMXB contribution to the total X-ray luminosity\,---\,are expected to be negligible. We checked the validity of this assumption by calculating the SFR for VCC1250, the bluest galaxy, and indeed found the corresponding HMXB contribution to $L^\mathrm{gal}_\mathrm{XRB}$ to be negligible, where the  SFR$_{V1250}$ $\approx$ 0.08 $M_{\odot}$ yr$^{-1}$ contributed $<$ 8\% of the total luminosity expected from XRBs). We thus ignore the $\beta$ term in Equation~\ref{eq:1} for the early-type galaxies. 
In general, the early-type galaxies are found to contain a larger portion of the total light within the central 2$\arcsec$ than the late-type sample, and thus have higher expected $L^n_\mathrm{XRB}$, with fractions ranging from 1.2--45.3\%, and 0.5--7 keV $L^n_\mathrm{XRB}$ values between $7 \times 10^{35}$--$1.7 \times 10^{38}\,\mathrm{erg\,s^{-1}}$.

\par We follow the same methodology as the late-types to estimate (i) the number of expected central XRBs and (ii) the probabilities that the 6 early-type detections are SMBHs. For our non-detections we define $N$ as the number of expected central XRBs with luminosities greater than or equal to the average sensitivity threshold ($L$ $\ge$ 3$\times$10$^{38}$ ergs s$^{-1}$; see \citealt{Gallo2010}) and for the 6 detections we define $N$ as the number of expected central XRBs with luminosities greater than or equal to the detected central X-ray luminosity ($L$ $\ge$ $L_{X}$).  Here, we assume the nuclear XRB luminosity is solely due to a distribution that follows the XLF of LMXBs \cite{Gilfanov2004}: $dN/dL$ = K$L^{-\alpha}$, where the slope for our given luminosity range is $\alpha$ = 1.8 and the normalization constant K is derived from the expected nuclear luminosity of XRBs. We find that the most likely number of total, central, XRBs in the early-type sample is projected to be consistent with 1, with probabilities $P_\mathrm{SMBH}$ for the six detections ranging from 90.1 to 92.7\% (see Table \ref{tab:4}). \\

We combine the results from the late- and early-type XRB contamination analysis.  The total number of nuclear XRBs follows a Poisson distribution, for which we find a most likely value of 3. The likelihood of detecting at least 11 SMBHs is $<$ 3\%, and at the 3$\sigma$ (99.7\%) confidence level 3 of our 11 detections are SMBHs.
%


\section{Active Fraction} \label{sec:results}
\begin{figure}[t]
 \center
\includegraphics[width = 8.4cm]{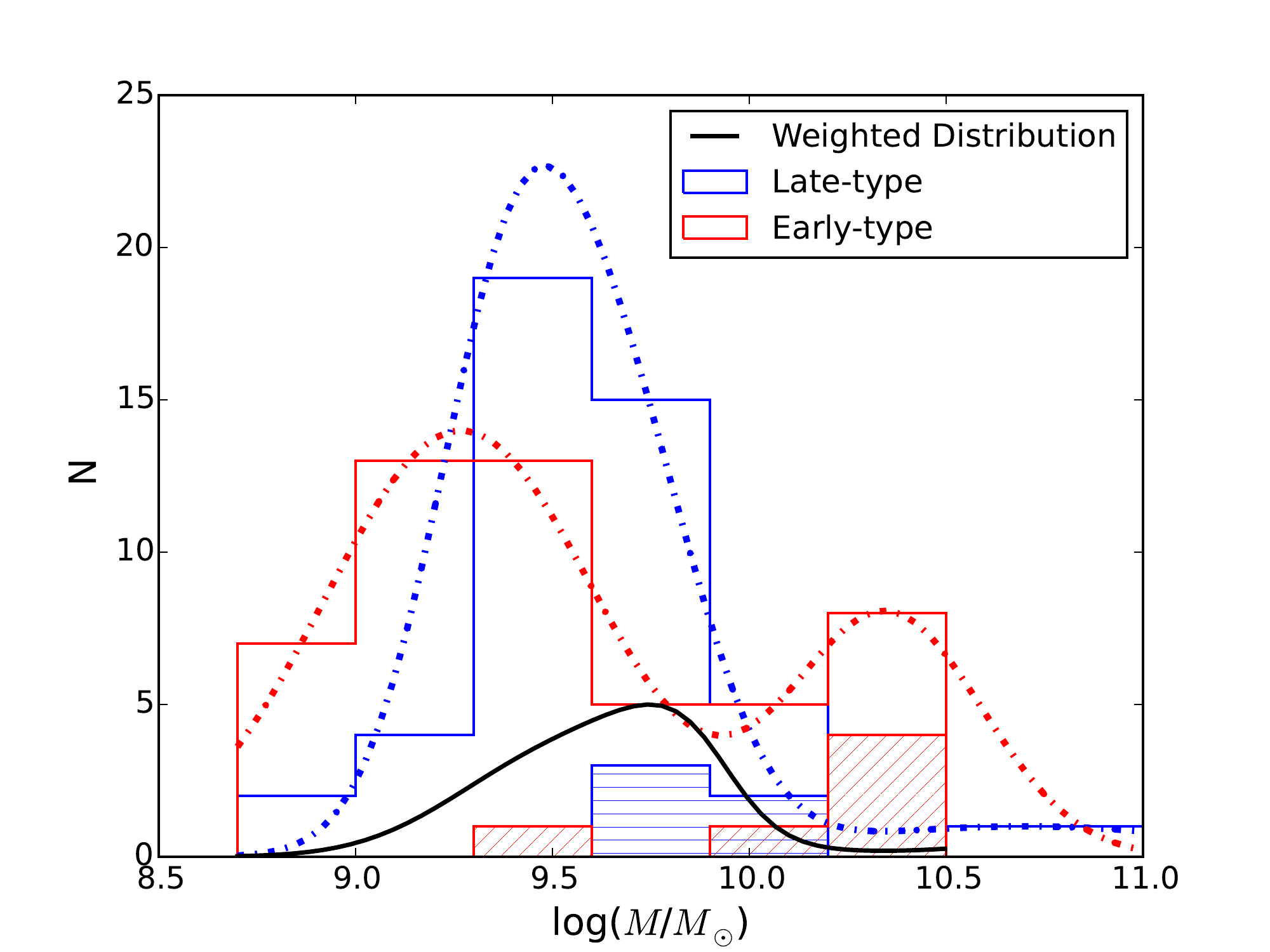}
\caption{Stellar mass distributions of the late- (open blue histogram) and early-type (open red histogram) samples. The distribution of galaxies with central X-ray detections are shown for both as histograms with diagonal (early-type) and horizontal (late-type) hatching.  Each sample has been fit with multiple Gaussians. The weighted distribution, defined as the ratio of the late-type distribution to the early-type distribution, is shown as a black curve. This is used to draw a sub-sample of the early-type galaxies with the same number and mass distribution as the late-type sample, allowing for a proper comparison of the active fraction between different morphological types.}
\label{fig:massweight}
\end{figure}

The probability-weighted X-ray detections discussed above can be employed to estimate the active fraction of nearby nucleated galaxies, down to a uniform X-ray luminosity threshold, and as a function of morphological class. Before comparing the active fractions between the late-type and early-type samples, we determine whether the two distributions are consistent with being drawn from the same mass distributions by using a weighting function. We follow the procedures outlined in \cite{Miller2012b}, and \cite{Balda2014}: the two samples are represented as a sum of Gaussians and we use a weighting function, defined as the ratio of the late-type distribution to the early-type distribution, to draw a subsample from the early-type distribution that has the same number of galaxies and mass distribution as the late-type sample (see Figure~\ref{fig:massweight}). We draw 500 such subsamples and find that, on average, they contain 5.5 objects with nuclear X-ray point sources, corresponding to 10.8\%$^{+11.3}_{-6.3}$. For the late-type sample, we found five our of 47 objects objects to have nuclear X-ray detections, corresponding to 10.6$\%$$^{+11.9}_{-4.9}$. Poisson statistics show that for an expected value of five active nuclei, there is a 38$\%$ chance of finding six or more active nuclei in a sample of 47. This argues for no statistically significant difference in the nucleation fraction of the early- and late-type samples. Across both morphological types, we find the active fraction to be 11.2$\%^{+7.4}_{-4.9}$.

\subsection{Regression Analysis}

\par In the following we explore a possible relation between the galaxy stellar mass, $M_{\ast}$, and the central X-ray luminosity, $L_{X}$.  Values for $M_{\ast}$ are derived from the galaxy's absolute luminosity ($L_{B}$), thus the $M_{\ast}$--$L_{X}$ relation is similar to the $L_{B}$--$L_{X}$ relation, which has been seen in nearby early-type galaxies \citep{Pellegrini2010}. As well, a relation between $M_{\ast}$--$L_{X}$ can be reflective of a possible relation between $M_{BH}$--$L_{X}$ due to the correlation between $M_{\ast}$ and $M_{BH}$. Similar analyses have been carried out for the AMUSE surveys, where a statistically significant correlation has been found between nuclear X-ray luminosity and host stellar mass (\citealt{Miller2012a, Miller2012b}). For the AMUSE-Field galaxies, \cite{Miller2012a} find a slope in the $M_{\ast}$--$L_{X}$ plane of $\beta$ = 0.71$\pm$0.10, which consistent with a non-zero value at a 3-$\sigma$ level. The tendency for the X-ray luminosity to increase less rapidly than $M_{\ast}$ can be interpreted as lower-mass galaxies being more X-ray luminous per unit stellar mass, however the large scatter ($\sim$0.7) allows for alternative explanations.
\par
\begin{figure}[t]
\center
\includegraphics[width = 8.4cm]{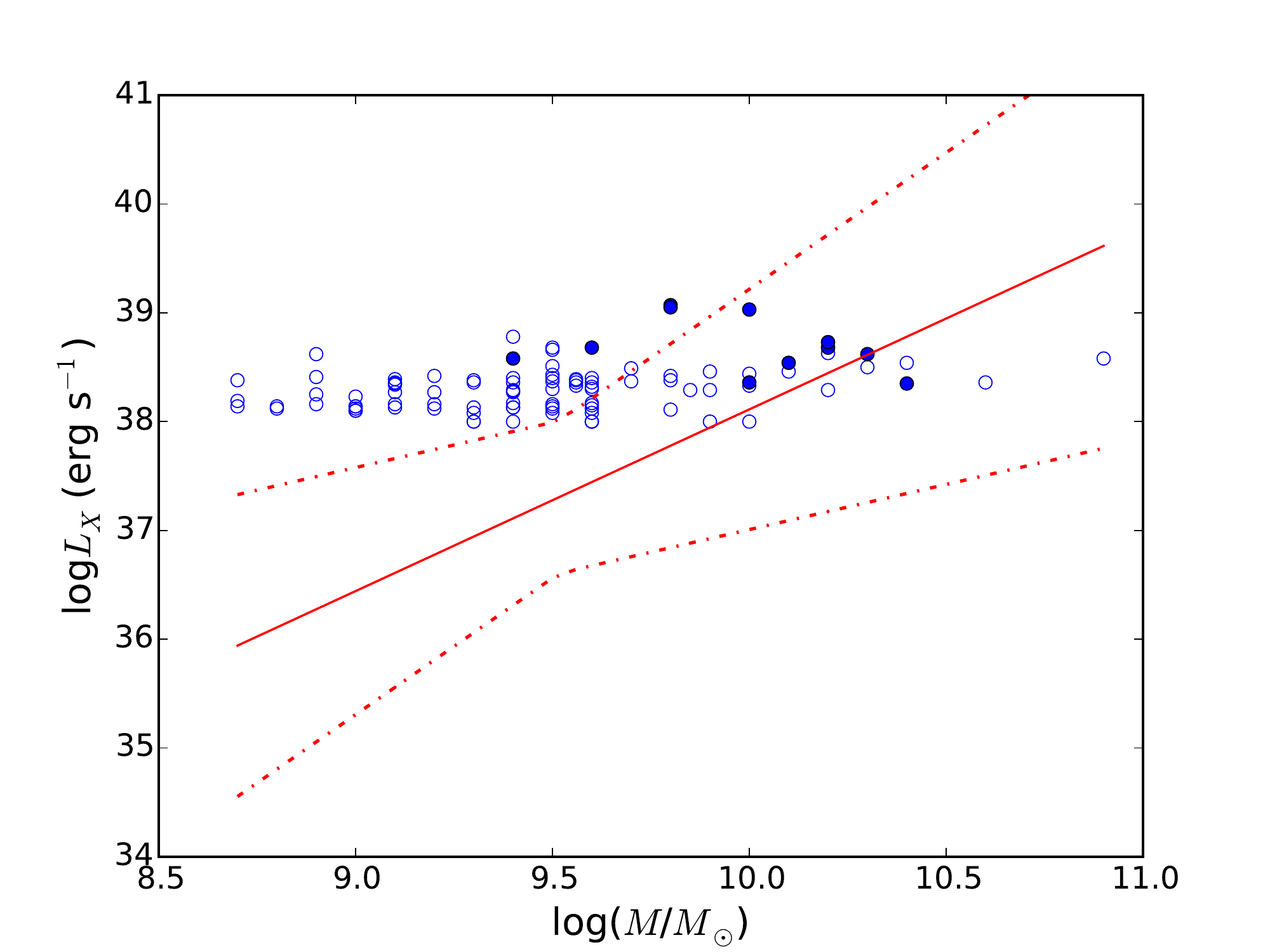}
\caption{Measured nuclear X-ray luminosities, $L_{\rm X}$, as a function of host galaxy stellar mass, $M_{\ast}$, for our full sample of late and early-type galaxies (98 systems). Filled circles mark detections and open circles mark upper limits. Error bars are taken to be 0.1 dex for both $L_{\rm X}$ and $M_{\ast}$.  The ``best-fitting relation" from our Bayesian linear regression analysis is shown as a solid red curve, with the dotted lines corresponding to 2$\sigma$ error bars. Here, 1$\sigma$ error bars correspond to the 16th and 84th percentiles of the posterior distributions.}
\label{fig:bigplot}
\end{figure}

 Motivated by the results of the AMUSE surveys, we assess the presence of a quantitative relation of the form:

\begin{equation}
\mbox{log} L_{X} - 38.2 = \alpha + \beta \times (\mbox{log} M_{*} - 9.52),
\label{eq:Kelly}
\end{equation}
where the variables are centered on the median values of the measured nuclear X-ray luminosity and host stellar mass for our combined sample of early and late-type nucleated systems. We use the Bayesian linear regression code ${\tt linmix\_err.pro}$, by \cite{Kelly2007}, which incorporates both uncertainties and censoring to determine the best-fit parameters. Similar to \cite{Miller2012a}, our errors are taken to be 0.1 dex on log($L_{\rm X}$), associated with the uncertainty in the distance. Our errors are taken to be 0.18 on log($M_{\ast}$); this is derived by adding in quadrature the error of 0.15 dex found in \cite{Seth2008} and the scatter of 0.1 dex in $M/L$ as presented in \cite{Bell2003}.
First, we treat every X-ray detected nucleus as indicative of an accreting SMBH, regardless of XRB contamination (i.e., we take $P_\mathrm{SMBH}$=100\% for all of the detected nuclei). 
To ensure uniform censoring, we set all of the upper limits below log($L_{\rm X}$) $=$ 38.0 (from deeper, archival exposures) to a value of 38. We find the median values of the intercept, slope, scatter and Pearson correlation coefficient to be: $\alpha$ = $-$1.50 $\pm$ 0.35, $\beta$ = 1.67 $\pm$ 0.42, $\sigma$ = 0.90 $\pm$ 0.20, and $r$ = 0.62 $\pm$ 0.11 (median values are taken from 10,000 draws of the posterior distributions, and the 1$\sigma$ errors quoted are calculated as the 16th and 84th percentiles). Figure~\ref{fig:bigplot} shows the distribution of log$L_{\rm X}$ vs. log$M_{\ast}$ for our complete sample with the ``best fit" line shown in red. Although the intrinsic scatter is large, the Pearson correlation value of 0.6 suggests a statistically significant correlation ($r$ = 0.6 for 98 points corresponds to a p value $<$ 0.01 given the sample size).

\par
\begin{figure*}[h]
\center
\includegraphics[width = 8.2cm]{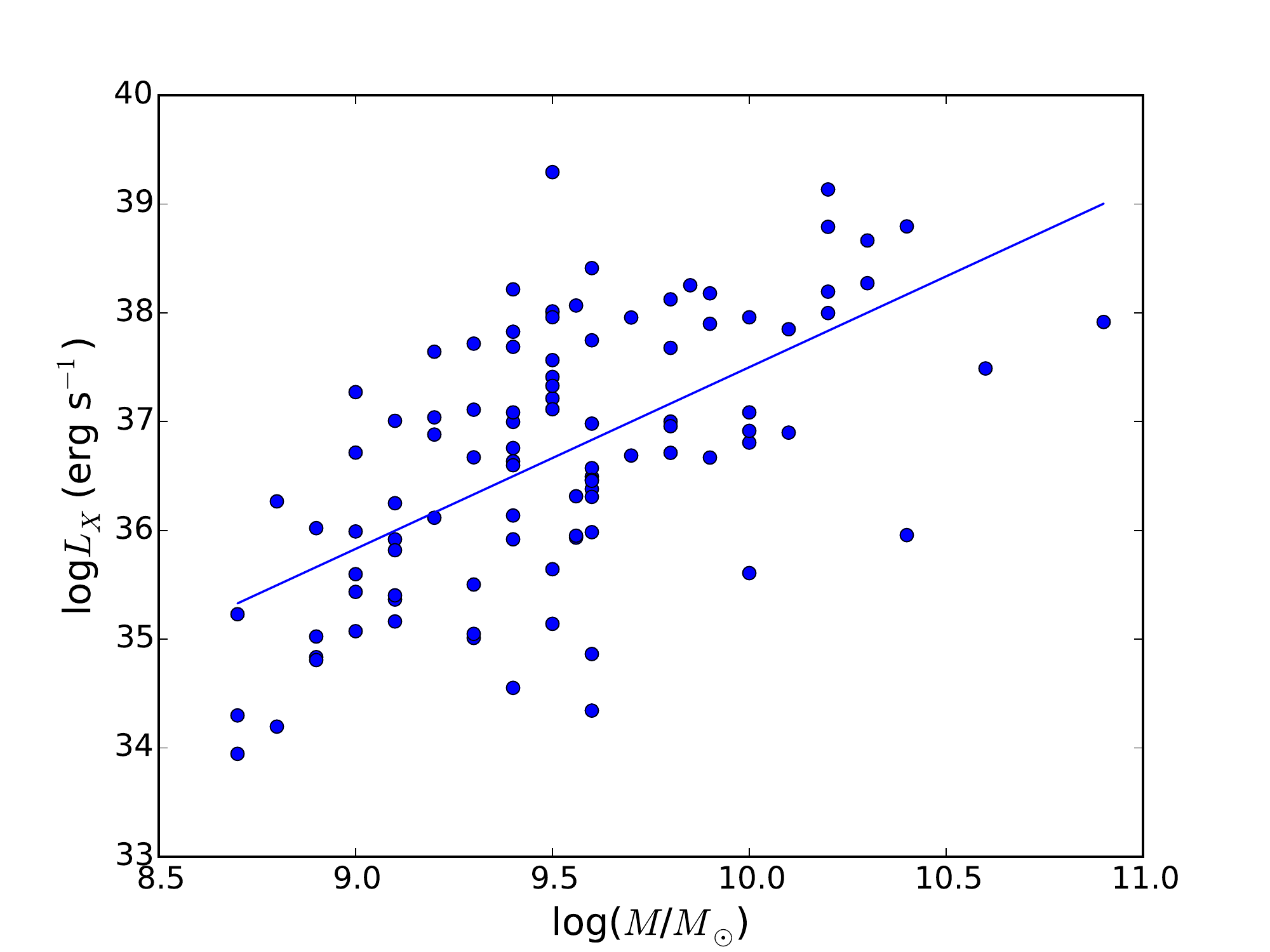}
\includegraphics[width = 8.2cm]{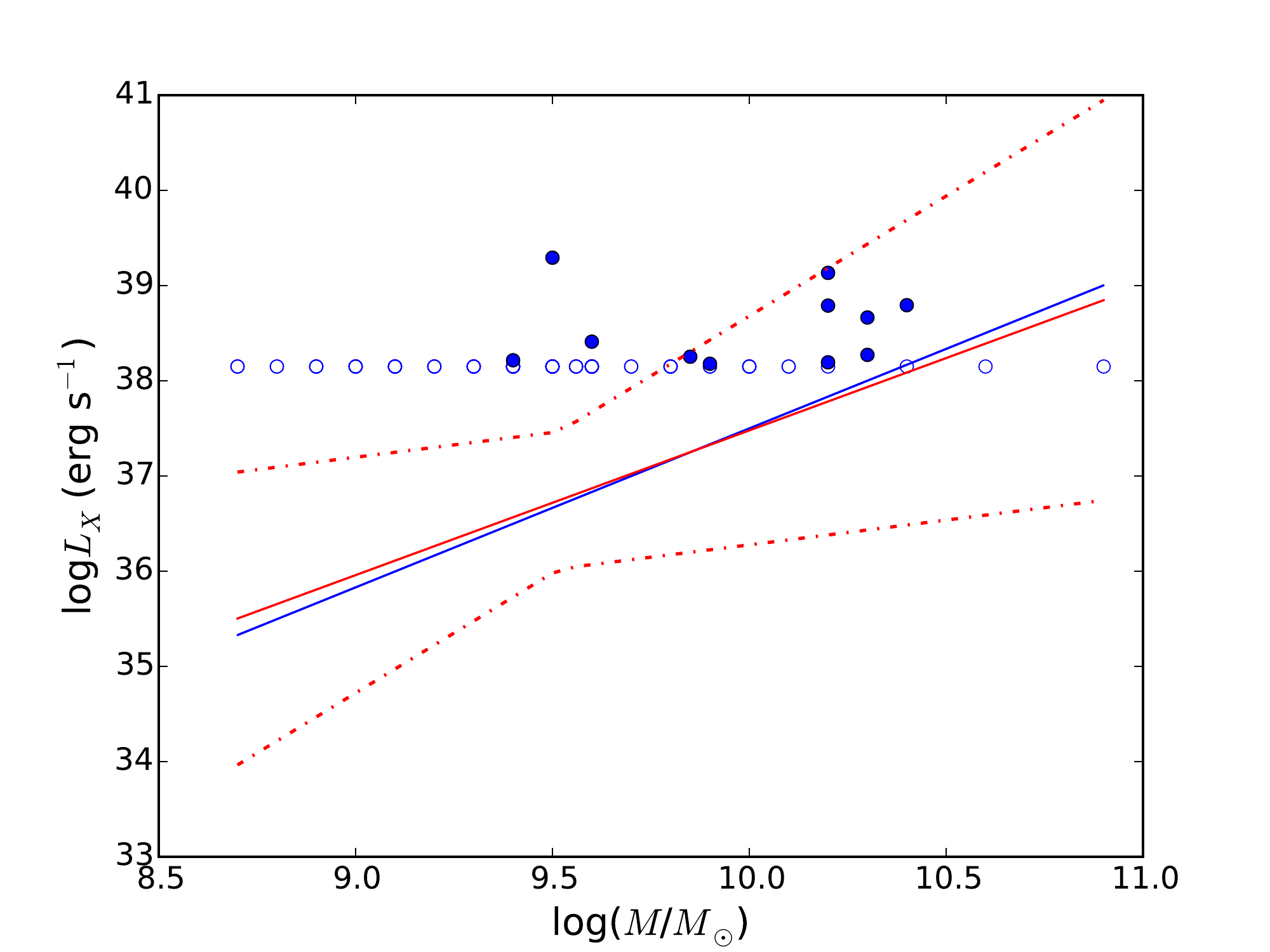}
\caption{{\it Left}: Starting from the same stellar mass distribution as or nucleated galaxy sample, we simulate a random distribution of $L_{\rm X}$ that follows the relation: log$L_{\rm X}$ $-$ 38.2 = $-$1.50 $+$ 1.67 $\times$ (log$M_{\ast}$ $-$ 9.52) (shown as a solid blue line), with an intrinsic, uniform scatter of 0.9, and Pearson correlation value of 0.62. {\it Right}: To match the actual nuclear X-ray detection fraction of our sample (i.e., 11\%), a luminosity detection threshold of log($L_{\rm X}$)=38.15 is chosen, so that all data points below that value turn into upper limits. The solid red line represents the ``best-fitting" relation recovered by  ${\tt linmix\_err.pro}$ (with dotted lines corresponding to 2$\sigma$ error bars), to be compared to the ``true" relation (blue solid line).  As also shown in Figure~\ref{fig:simpostD}, the median value of the slope posterior distribution for the 11\% detection fraction data set is consistent with the ``true" slope to within 2$\sigma$.}
\label{fig:simulate}
\end{figure*}

\begin{figure*}[h]
\center
\includegraphics[width = 16.4cm]{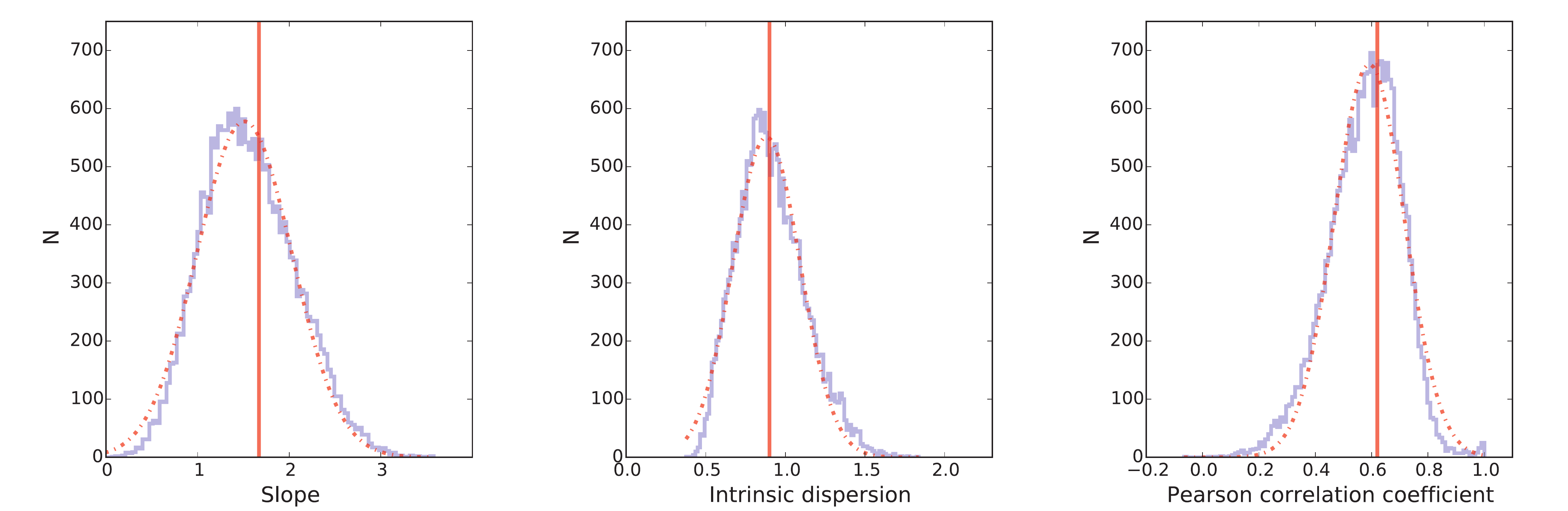}
\caption{The posterior distributions of the slope, intrinsic scatter, and the linear correlation coefficient for the simulated data points shown in the right panel of Figure~\ref{fig:simulate} (11\% detection fraction) are plotted here in purple, and fit with Gaussian distributions (red dashed curves). For comparison, the $true$ values of the slope, scatter and correlation coefficient, corresponding to the simulated data in the left panel of Figure~\ref{fig:simulate} are shown as vertical red lines. The ``true" values are consistent with the medians of the purple posterior distributions at the 2$\sigma$ level, indicating that the regression analysis is robust even for detection fractions as low as $\sim$ 10\%.}
\label{fig:simpostD}
\end{figure*}

 A valid concern, however, has to do with the ability of {\tt linmix\_err.pro} to recover an intrinsic correlation when dealing with a large fraction of upper limits, rather than detections (i.e., 89\% of the data points, in our case). In order to assess this, following \cite{Kelly2007} we start from the same stellar mass distribution as our data and simulate a random distribution of $L_{\rm X}$ that follows the above mentioned relation: log$L_{\rm X}$ $-$ 38.2 = $-$1.50 $+$ 1.67 $\times$ (logM$_{\ast}$ $-$ 9.52), with an intrinsic scatter of 0.9, and Pearson correlation value of 0.62, as shown in the left panel of Figure~\ref{fig:simulate}.
We then run {\tt linmix\_err.pro} on the simulated points, adopting progressively higher X-ray luminosity detection thresholds values to lower the detection fraction, down to the limiting case of only 11\% of the data resulting into detections, shown in the right panel of Figure~\ref{fig:simulate}. 
For this case, Figure~\ref{fig:simpostD} shows the full posterior distributions of $\beta$, $\sigma$, and $r$ recovered by {\tt linmix\_err.pro}, compared to the ``true" values, indicated by the solid red vertical line. In spite of the very high fraction of upper limits, the median values of the parameters' posterior distributions agree within 2$\sigma$ with the ``true" values. 

\par Finally, in order to incorporate the results of our XRB contamination assessment into the correlation analysis we follow the probabilistic method described by  \cite{Alfvin2016}, i.e., we run {\tt linmix\_err.pro} 200 times, probabilistically varying whether each X-ray nucleus is treated as an upper limit or a detection according to the probabilities estimated in the previous Section (see Table~\ref{tab:4}). We find that the distribution of slopes is consistent with $\beta$ = 1.67 $\pm$ 0.42, but the uncertainties do not rule out a zero slope.
 
As a consistency check on the fitting, we stacked the data for late-type galaxies, using the de-flared data in the $0.5$--$7.0$~keV band and applying the diffuse-gas corrections where appropriate. Since not all of the galaxies are
astrometrically corrected, we use a detection cell $5^{\prime\prime}$ across instead of the $2^{\prime\prime}$ cell quoted above. We adopted bins of $9.0 \le \log M_* < 9.5$, $9.5 \le \log M_* < 10.0$, and $10.0 \le \log M_* < 10.5$,
in which we measure $L_X = 0.2\pm0.1 \times 10^{38}$\,erg\,s$^{-1}$,  $L_X = 1.0\pm0.2\times 10^{38}$\,erg\,s$^{-1}$, and $L_X = 3.5\pm0.6 \times 10^{38}$\,erg\,s$^{-1}$, respectively. These errors incorporate the systematic error in $L_X$ and $M_*$.

Assuming that black holes follow a $L_X$--$M_*$ relation with some scatter, which is the same assumption used in the fitting above, we find that the three bins rule out a non-zero slope at greater than $3\sigma$. The high-mass bin only
has several galaxies, so the effect of intrinsic scatter may be severe. However, if we ignore it, the remaining two average luminosities are still inconsistent at greater than 3$\sigma$. There is also a large variance in the  individual exposure times, but when omitting the longest exposures a non-zero slope is ruled out at greater than $3\sigma$. 

We examined the effect of XRB contamination on the average luminosities following the procedure outlined above to assess how many counts are likely from undetected XRBs. The effect is most pronounced in the lowest mass bin, where there are only nine total counts and we expect at least this many about 30\% of the time. Indeed, the luminosity in this bin is consistent with that expected from XRBs, since for the exposure-weighted average mass in this bin we expect $L_X < 2\times 10^{37}$\,erg\,s$^{-1}$. In contrast, there are only a few galaxies in the high-mass bin and there are two strong detections with low odds of being XRBs (see above). Even if one of these detections is an XRB, the residual luminosity is greater than $3\sigma$ above that in the low-mass bin. 

Compared to the regression analysis above, stacking uses more information about the number of counts above background but sacrifices information about the stellar mass distribution. The consistency of these approaches in finding a non-zero $L_X - M_*$ slope implies that this result is real and not an artifact of a particular scheme.

\subsection{Nuclear star cluster compactness}
\begin{figure}[t]
 \center
\includegraphics[width = 8.4cm]{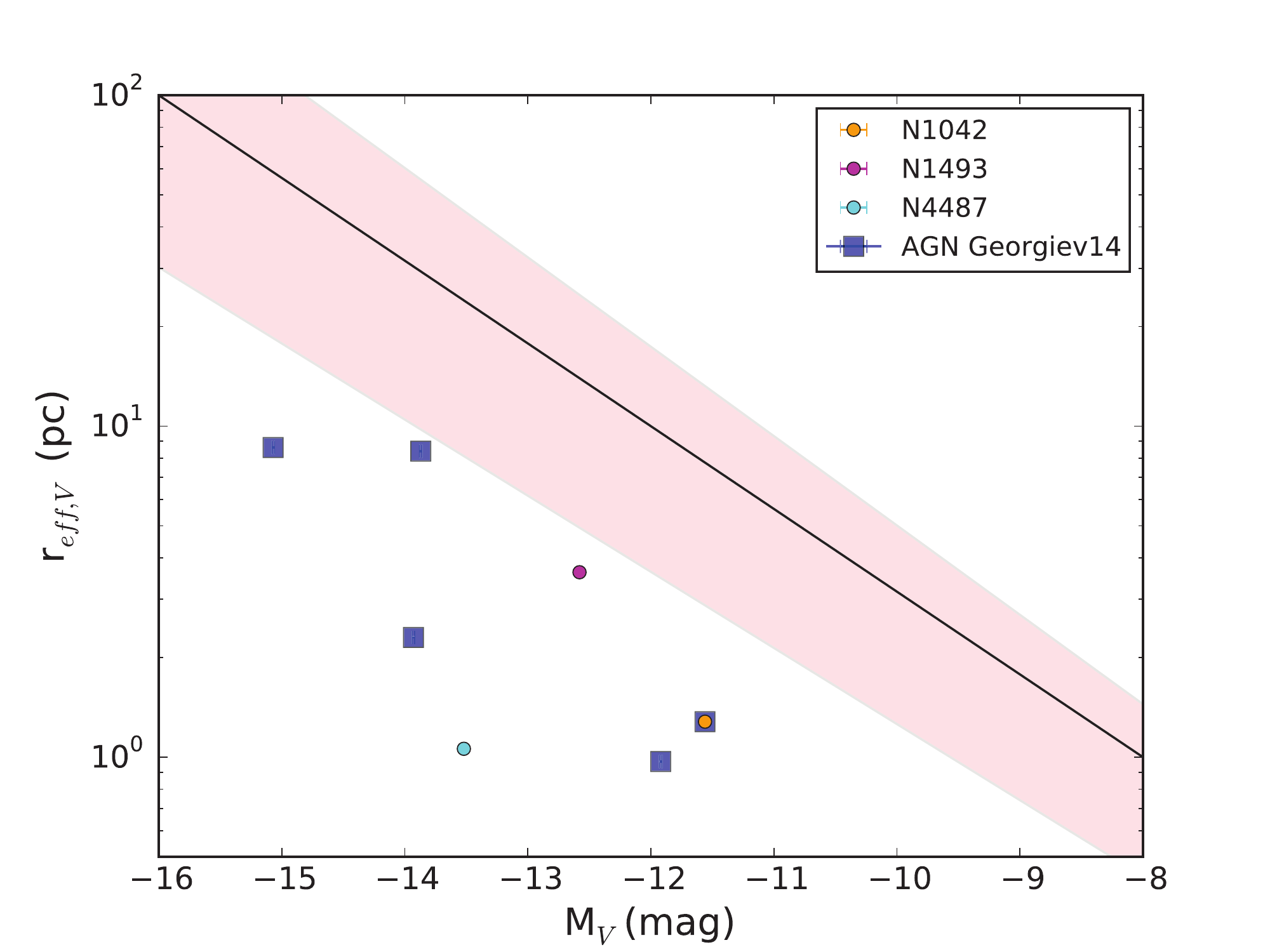}
\caption{Size-luminosity relation for the NSCs in the catalogue of spiral galaxies analyzed in \cite{Georgiev2014} is shown in black, with region of error shown in red: log $r_{eff}$ = -2.0$\pm$0.2 -- 0.25$\pm$0.01 $M_{V}$. Blue squares denote the galaxies with known AGN in \cite{Georgiev2014}. We add three late-type detections from our sample that are included in their catalogue: NGC 1042 (orange circle with blue square), NGC 1493 (pink circle) and NGC 4487 (cyan circle).  NGC 1042 was recognized as an AGN by \cite{Georgiev2014}  but was not included in their compactness study due to S/N $<$ 30 (NGC 1493 and NGC 4487 were excluded from their parent sample for similar reasons). Error bars on $M_{V}$ are $<$ 0.1 mag. We also find that our nucleated galaxies with AGN tend to have more compact effective radii at a given luminosity.}
\label{fig:geor}
\end{figure}

One of the most recent, comprehensive studies on nucleated late-type galaxies is presented in \cite{Georgiev2014}, where 224 spiral galaxies with NSCs were thoroughly analyzed via {\it HST} observations. They find that all of the well-resolved (S/N $>$ 30) NSCs with known AGN in their sample appear more compact than the parent nucleated sample (see figure 12 in their paper).  They interpret this as compact, excess flux in the F606W filter due to H$\alpha$ emission produced by weak AGN activity and/or star formation.
34 of our 47 late-type galaxies were included in their study, including 3 (NGC 1042, NGC 1493, and NGC 4487) of the 5 objects that we identified as active via X-ray analyses.  One of the objects, NGC 1042, was recognized as an AGN by \cite{Georgiev2014} based on results from \cite{Shields2008}, but was not included in their compactness study due to S/N $<$ 30 (NGC 1493 and NGC 4487 were excluded from their parent sample for similar reasons). Here, we extend their analysis by plotting the 3 objects on the same size-luminosity plot (r$_{eff}$ vs. M$_{V}$) as in \cite{Georgiev2014} (see Figure~\ref{fig:geor}). All 3 objects lie below the best-fit relation found in \cite{Georgiev2014}.  We confirm the trend that nucleated galaxies with AGN tend to have more compact effective radii at a given luminosity, possibly caused by the presence of a weak AGN and/or a younger stellar population that is more concentrated than the surrounding NSC stars.


\section{Summary and Discussion}
In this work, we report on  {\it Chandra} observations of a distance-limited sample of 98 galaxies known to harbor NSCs, with the aim to characterize the incidence and intensity of the AGN (or rather, weakly accreting SMBH) population down to a uniform X-ray detection threshold of $10^{38}$ erg s$^{-1}$, i.e. comparable to Eddington limit for a stellar mass object. 
From a theoretical standpoint, a somewhat higher active fraction can be expected in nucleated galaxies (i.e., compared to non-nucleated galaxies over the same mass range), due to the NSC facilitating either the {\it formation} of a nuclear SMBH, its {\it feeding}, or a combination of the two. Naively assuming a roughly constant mass accretion rate onto the SMBH, the former scenario would result into a higher fraction of active galaxies among nucleated vs. non-nucleated. At the same time, the latter scenario, with enhanced fueling, can be expected to be more relevant for nucleated late-types, home to bluer NSCs and thus enhanced gas reservoir for black hole accretion (\citealt{Antonini2015,Naiman2015,Gnedin2014,Muratov2010}; see, however, \citealt{Antonini2015}).

The data set presented here is comprised of new {\it Chandra} data for 47 late-types, plus archival data for 51 early-types (\citealt{Gallo2010,Miller2012a,Miller2012b}), enabling us to (i) improve on previous measurements of the nucleated galaxy active fraction (e.g., \citealt{Seth2008}) by adopting a uniform criterion for identifying and measuring accretion-powered emission; (ii) carry out a morphology-dependent analysis, which in turn could break the degeneracy between an intrinsically higher fraction of nucleated galaxies actually hosting SMBHs vs.\ an enhancement in the fueling of such SMBHs, likely resulting into a higher active fraction for the late-type sample.  

\begin{figure}[]
\center
\includegraphics[width = 8.4cm]{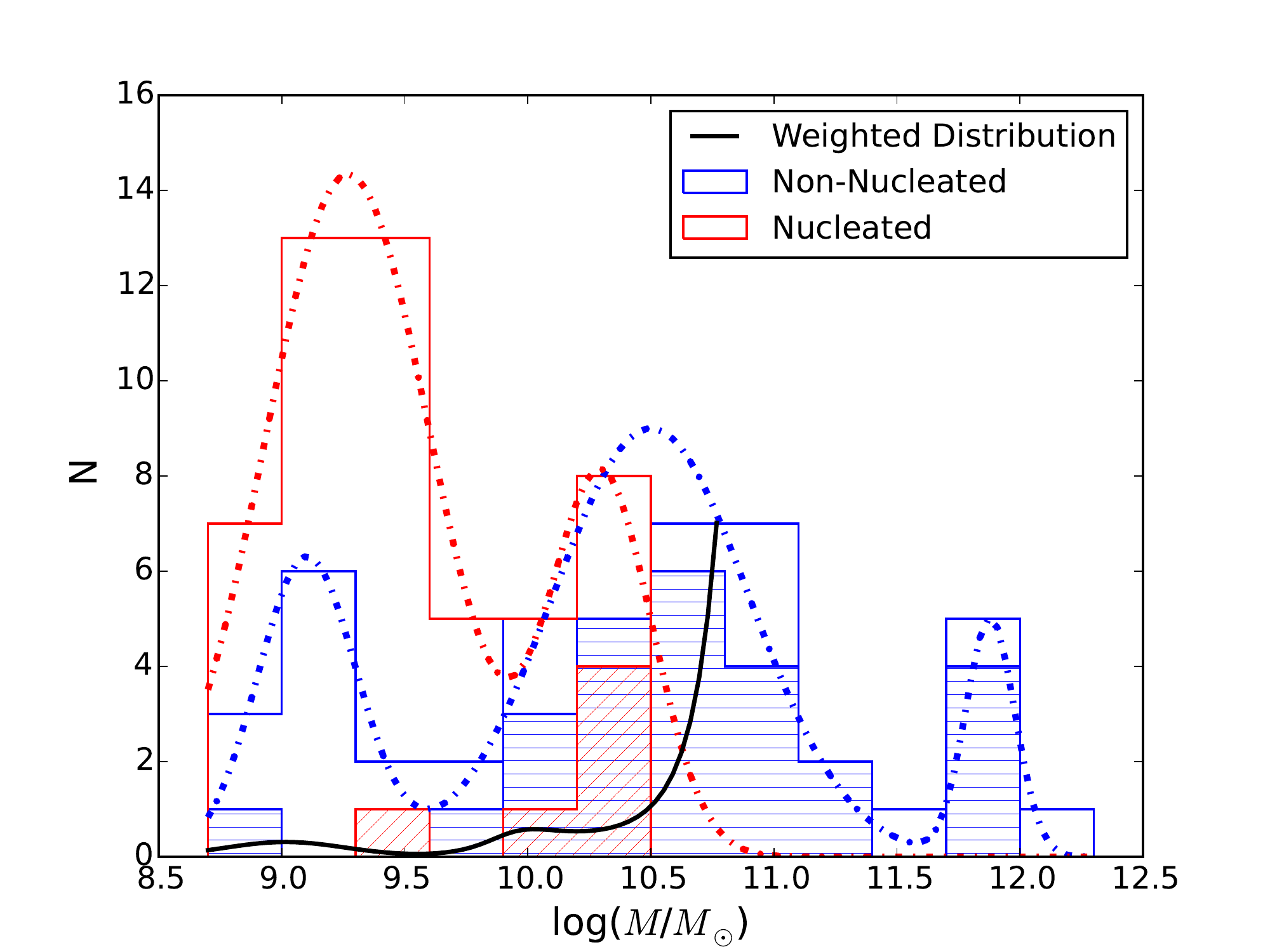}
\caption{Same as Figure~\ref{fig:massweight}, the 100 Virgo early-types targeted as part of the Virgo Cluster Survey, 51 of which are nucleated (red open histogram) while 49 are not (blue open histogram).  The distribution of galaxies with central X-ray detections are shown for both as histograms with diagonal (nucleated) and horizontal (non-nucleated) hatching.  Each sample has been fit with multiple Gaussians. The weighted distribution, defined as the ratio of the non-nucleated distribution to the nucleated distribution, is shown as a black curve. This is used to draw a sub-sample of nucleated galaxies with the same number and mass distribution of the non-nucleated sample within the mass range 8.7 $<$ log(M/M$_{\odot}$) $<$ 10.5, allowing for a proper comparison of the active fraction between nucleated and non-nucleated systems.}
\label{fig:massweightvirgo}
\end{figure}

With respect to point (i), after accounting for X-ray binary contamination to the nuclear X-ray signal,  we estimate an active fraction of 11.2$\%^{+5.0}_{-7.3}$ across the whole sample. This is in broad agreement with previous estimates by \cite{Seth2008}. A more detailed comparison with their results is warranted in order to compare the diagnostics power of high resolution X-ray imaging observations vs.\ different wavelengths/methods. Our sample of 98 objects represents a distance-limited sub-sample drawn from the original \cite{Seth2008} sample; for these 98 nucleated galaxies, they present uniform optical diagnostics, in the form of BPT optical emission line ratios, for 53 objects.  Only 2 out of 53 (i.e., NGC 5879 and NGC 4411B) are identified as active (with six more, namely NGC 1042, NGC 3423, NGC 4206, NGC 4517, VCC 1619, NGC 4625 identified as composite). By comparison, we identify 7 out 53 as likely AGN (NGC 1042\footnote{\cite{Shields2008} report on follow-up optical spectroscopy of NGC 1042 using the TWIN spectrograph at the Calar Alto 3.5 m telescope (with 3" wide extraction region for consistency with the SDSS fiber aperture), and conclude that the source is indeed an AGN.}, NGC 5879, VCC 784, VCC 1250, VCC 1283, VCC 1619, and VCC 1883; notice, however, that we do not detect NGC 4411B in X-rays). 
While recognizing that our assessment of X-ray binary contamination does not take into account potential differences in the X-ray binary XLF between nucleated and non-nucleated galaxies, this direct comparison suggests that in the nearby universe, {\it Chandra} observations are superior to standard emission line ratio diagnostics to identify low-level SMBH accretion.
We close this comparison by noting that the 10 per cent active fraction quoted by \cite{Seth2008} stems from a sample of 75+ galaxies with (non-uniform) spectral coverage. Additional AGN were identified through archival X-ray observations, with {\it ROSAT, XMM-Newton, Chandra}, as well as radio observations. Overall, the two galaxies that were identified by Seth et al. as actively accreting via X-ray diagnostics have stellar masses slightly below the mean mass of the 7 optical galaxies found to host AGN via BPT diagnostics (i.e., log($M/M_{\odot}$)= 10.3 $\&$ 10.2, vs. log($M_{\rm mean}/M_{\odot}$) = 10.4).

With respect to point (ii), as discussed in Section 4, after accounting for the different mass distributions, we find no statistically significant difference in the active fraction among nucleated late and early-types (10.8$\%$ $^{+11.3}_{-6.3}$ vs. 10.6$\%$ $^{+11.9}_{-4.9}$, respectively).  

Recently, \cite{Antonini2015} employed semi-analytical models to make predictions about the rate of coexistence of NSCs and SMBHs in local galaxies, regardless of activity level.  Among the simulated nucleated galaxies with stellar mass $<$ 10$^{10}M_{\odot}$, the early-types are predicted to have a larger SMBH {\it occupation} fraction than late-types (the reader is referred to figure 7 in their paper).  Assuming the occupation model presented in  \cite{Antonini2015}  is correct, the similar AGN fractions observed in early- vs. late-type samples suggests that late type galaxies may more efficiently fuel their SMBHs.

The overall active fraction for our nucleated sample is also consistent with the well known 10\%  active fraction often quoted for the general population, albeit across a much larger dynamical range in galaxy mass, redshift, as well as accretion-luminosity sensitivity (e.g,  \citealt{Ho1997,Kewley2006}).
However, in the light of the well known trend of increasing apparent active fraction with host stellar mass (due to the ability to probe progressively lower Eddington ratios towards higher masses), a more meaningful comparison is one between nucleated vs.\ non-nucleated galaxies across comparable stellar (and hence, black hole) mass ranges, and down to a uniform accretion luminosity threshold. This can be accomplished by looking at the 100 early-type galaxies that comprise the ACS Virgo cluster survey (51 nucleated, 49 non-nucleated), and for which we also have uniform {\it Chandra} coverage (by comparison, we lack a parent sample of non-nucleated late-type with uniform {\it Chandra} coverage).  On average, the non-nucleated sample has a stellar mass of log($M$/$M_{\odot}$) = 10.3 that spans between 8.7 $<$ log(M/M$_{\odot}$) $<$ 12.3, while the nucleated sample has a lower average stellar mass of log($M$/$M_{\odot}$) = 9.5 that ranges between 8.7 $<$ log(M/M$_{\odot}$) $<$ 10.5.  Thus, we follow the same approach as described in Section~\ref{sec:results} in order to carry out a mass-weighted comparison of the active fraction between the nucleated and non-nucleated Virgo galaxies within the mass range 8.7 $<$ log(M/M$_{\odot}$) $<$ 10.5 (see Figure~\ref{fig:massweightvirgo}). On average, the nucleated subsample has 4 objects with nuclear X-ray point sources, corresponding to an active fraction of 15\%$^{+20}_{-10}$. The non-nucleated subsample has 10 objects with nuclear X-ray point sources, corresponding to an active fraction of 38\%$^{+38}_{-18}$. Poisson statistics show that for an expected value of 10 nucleated galaxies, there is a 3\% chance of finding 4 or fewer objects. Thus, the nucleated subsample is measured to have a marginally lower active fraction than the non-nucleated sample; this supports our conclusion that nucleated galaxies do not appear to have enhanced AGN fractions. 
\\
\par The main results and implications of this work can be summarized as follows: \\
1. After correcting for contamination to the nuclear X-ray signal from bright X-ray binaries and accounting for the different stellar mass distributions, we find no statistically significant difference in the active fraction of early- vs. late-type nucleated galaxies, with $f$=10.6$\%^{+11.9}_{-4.9}$ and 10.8$\%^{+11.3}_{-6.3}$. Across the whole sample, we measure an active fraction $f$=11.2$\%^{+7.4}_{-4.9}$ (1$\sigma$ C.L.), in agreement with previous estimates by \cite{Seth2008}.  For the early-type nucleated galaxies we carry out a controlled comparison with a parent sample of non-nucleated galaxies and find no statistically significant difference in the active fraction. We conclude that nucleated galaxies do not appear to have enhanced AGN fractions. \\
2. We investigate a relationship between the host galaxy stellar mass, $M_{\ast}$, and the central X-ray luminosity, $L_{X}$ of the form: $\mbox{log} L_{X} - 38.2 = \alpha + \beta \times (\mbox{log} M_{*} - 9.52)$. We find the most-likely values for the y-intercept, slope, intrinsic scatter, and Pearson correlation coefficient to be: $\alpha$ = -1.50 $\pm$ 0.35, $\beta$ = 1.67 $\pm$ 0.42,  $\sigma$ = 0.90 $\pm$ 0.20, and $r$ = 0.62 $\pm$ 0.11.  When we incorporate XRB contamination into our analysis, no statistically significant relation is found between the contamination-weighted X-ray luminosities and inferred host stellar masses.\\
3. We extend the size-luminosity analysis analyzed in \cite{Georgiev2014} by adding three late-type detections from our sample that are included in their catalogue: NGC 1042, NGC 1493, and NGC 4487. We confirm the general trend that nucleated galaxies with AGN tend to have more compact effective radii at a given luminosity, possibly caused by the presence of a weak AGN and/or a younger stellar population that is more concentrated than the surrounding NSC stars. 
\par We close by noting that, at the luminosity levels we are operating, a careful assessment of the X-ray binary contamination to the nuclear X-ray signal is crucial in order to draw meaningful conclusions. In the absence of a quantitative prescription for the XLF of NSCs, our methodology, which admittedly relies on XLFs established for the non-nuclear regions of nearby galaxies, may indeed underestimate such contribution. If this is the case, however, the active fractions quoted in this work are somewhat over-estimated, further strengthening our conclusion that the presence of a NSC does not favor nor enhance AGN activity. \\

We thank our referee, Anil Seth, for his constructive and thorough suggestions that improved this paper. Support for this work was provided by the National Aeronautics and Space Administration through Chandra Award Number GO5-16082X issued by the Chandra X-ray Observatory Center, which is operated by the Smithsonian Astrophysical Observatory for and on behalf of the National Aeronautics Space Administration under contract NAS8-03060.

\begin{table*}
	\begin{center}
	\caption{Nucleated late-type galaxy properties}
	\label{tab:1}
	\footnotesize
	\begin{tabular}{ llcccccccccccccc }
		\hline
		\hline  
		\multicolumn{1}{c}{Galaxy} & \multicolumn{1}{c}{obsID}  &  \multicolumn{1}{c}{Type} &\multicolumn{1}{c}{Exposure} & \multicolumn{1}{c}{Distance} & \multicolumn{1}{c}{log$M_{\ast}$} & \multicolumn{1}{c}{log$L_{\rm X}$}\\
	\multicolumn{1}{c}{} & \multicolumn{1}{c}{} &  \multicolumn{1}{c}{} & \multicolumn{1}{c}{(ks)}  & \multicolumn{1}{c}{(Mpc)} & \multicolumn{1}{c}{(log$M_{\odot}$)} & \multicolumn{1}{c}{(log(erg s$^{-1}$))}\\
	\multicolumn{1}{c}{(1)} & \multicolumn{1}{c}{(2)} & \multicolumn{1}{c}{(3)} & \multicolumn{1}{c}{(4)}  & \multicolumn{1}{c}{(5)} & \multicolumn{1}{c}{(6)} & \multicolumn{1}{c}{(7)} \\
	\hline
	NGC 300		& 16028 (A)	&	Scd 		&	63.84	& 	1.9		&	9.3	&	$<$ 35.4					\\
  	NGC 337a 	& 16976 (C16)	& 	SABd 	& 	3.20  	&	14.3		&	9.4	&	$<$ 38.36					\\
 	NGC 406   	& 16977 (C16)	& 	Sc  		& 	4.70	 	&	17.5	 	&	9.7  	&	$<$ 38.35					\\
	NGC 428   	& 16978 (C16)	& 	SABm 	& 	2.97	 	&	15.9		&	9.8 	&	$<$ 38.36					\\
	NGC 1042	& 12988 (A)	& 	SABc	& 	30.00	&	18.0		&	10.0 	&	38.34					\\
  	ESO 548-G29 	& 16979 (C16)	&	SABb	& 	3.70	 	&	16.2	 	&	9.56*&	$<$ 38.76					\\
  	ESO 418-8 	& 16980 (C16)	& 	SABd  	& 	3.12	 	&	14.1		&	9.4	&	$<$ 38.35		 			\\
	NGC 1483 	& 16981 (C16)	& 	SBbc	& 	2.26	 	&	12.6		& 	9.1 	&	$<$ 38.26 				\\
	NGC 1493	& 7145 (A)	&	SBc		&   	9.24		&	11.3	 	&	9.6	&	 38.68					\\
  	NGC 1688 	& 16982 (C16)	&	SBc		& 	2.03	 	&	13.4		& 	9.5 	&	$<$ 38.40					\\
  	NGC 1892 	& 16983 (C16)	&	Sc		& 	2.56	 	&	15.2		& 	9.6 	&	$<$ 38.37					\\
	NGC 2082	& 7838 (A)	&	SBb		&	5.07		&    	15.3 		& 	9.8 	&	$<$ 38.10					\\
  	NGC 2104 	& 16984 (C16)	&	SBm		&	2.63	 	& 	12.8		&	9.1	&	$<$ 38.35					\\
  	UGC 4499 	& 16985 (C16)	&	Sd		&	2.17	 	& 	12.6 		&	8.7 	&	$<$ 38.36					\\
	NGC 3423	& 16346 (A)	&	Sc		&	47.02 	&	14.7		&	9.9	&	$<$ 37.34					\\
  	NGC 3455 	& 16986 (C16)	&	SABb	&	4.35	 	& 	16.9		& 	9.5 	&	$<$ 38.35					\\	
  	NGC 3501 	& 16987 (C16)	&	Sc		&	4.55	 	& 	17.4		& 	9.56*&	$<$ 38.28					\\
	NGC 3782 	& 16988 (C16)	&	Scd		&	4.94	 	& 	13.6		&	9.56*& 	$<$ 38.37					\\
  	NGC 3906 	& 16989 (C16)	&	SBcd	&	4.50	 	& 	16.9		& 	9.3	&	$<$ 38.36					\\
	NGC 3913	& 7856 (A)	&	Scd		&	4.70		&	17.1 		& 	9.4 	&	$<$ 37.65					\\
  	NGC 3949 	& 16990 (C16)	&	Sbc		&	3.39	 	& 	14.6 		& 	9.9	&	$<$ 38.27					\\
	NGC 4030	& 11670 (A)	&	Sbc		&	14.86	&	21.1 		& 	10.9  &	$<$ 37.94					\\
  	NGC 4144 	& 16991 (C16)	&	SABc	&	1.05	 	& 	7.2		& 	9.0 	&	$<$ 38.22					\\
  	NGC 4183 	& 16992 (C16)	&	Sc	   	&	3.58	 	& 	16.2 		&	9.6	& 	$<$ 38.35					\\
	NGC 4204	& 7092 (A)	&	SBd		&	1.99		& 	13.9 		&	9.56*&	$<$ 38.49					\\
	NGC 4206 	& 16993 (C16)	&	Sbc		&	2.00	 	&	11.3		&	9.5 	&	$<$ 38.38					\\ 
	NGC 4299	& 7834 (A)	&	SABd	&	5.07		&	16.8		&	9.3 	&	$<$ 38.07					\\
	NGC 4411b	& 7840 (A)	&	SABc	&	4.89		&	19.1		&	9.6	& 	$<$ 38.31					\\
  	NGC 4487 	& 16994 (C16)	&	Sc		&	2.82	 	&	14.7		&	9.8	&	39.07			 		\\
  	NGC 4496a 	& 16995 (C16)	& 	SBd		&	2.72	 	& 	15.0 	 	& 	9.7	& 	$<$ 38.48					\\
  	NGC 4517 	& 16996 (C16)	&	Sc		& 	4.06	 	& 	16.5	 	&	10.6 	&	$<$ 38.34					\\
	NGC 4618	& 7147 (A)	&	SBm		&	9.11		&	10.7	 	&	9.6 	&	$<$ 37.52					\\
	NGC 4625	& 7098 (A)	&	SABm	&	7.10		&	11.7		&	9.4 	&	$<$ 38.27					\\
	NGC 4701	& 7148 (A)	&	Sc		&	9.04		&	11.1		&	9.3 	&	$<$ 37.56					\\
  	NGC 5023 	& 16997 (C16)	&	Sc		&	1.02	 	& 	5.4		&	8.7	&  	$<$ 38.18					\\
	NGC 5068	& 7149 (A)	&	Sc		&	6.67		&	8.7		&	10.0 &	$<$ 36.51					\\
  	UGC 8516	& 16998 (C16)	&	Sc		&	3.64	 	& 	16.7 		&	9.3 	&	$<$ 38.35			 		\\
	NGC 5585	& 7150 (A)	& 	SABc	&	5.32		&	10.5		& 	9.4 	&	$<$ 37.12					\\
	NGC 5879	& 2241 (A)	&	Sbc		&	88.77	&	15.0		&	10.0 	& 	38.55					\\
	NGC 6239 	& 16999 (C16)	&	SBb		&	4.25		& 	16.9 		&	9.6 	&	$<$ 38.28					\\
	IC 5052 		& 17000 (C16)	&	SBcd	&	1.02	 	& 	6.0 		&	9.1	&	$<$ 38.10					\\
  	IC 5256 		& 17001 (C16)	&	SBd  	&	1.39	 	& 	9.9 		& 	9.56*&	$<$ 38.32					\\
	NGC 7424	& 3495 (A) 	& 	Sc		&	23.24 	&	10.8		& 	9.6 	&	$<$ 37.11					\\
	NGC 7690 	& 17002 (C16)	&	Sb		&	4.43	 	&	18.2		& 	9.8	& 	39.05					\\
	UGC 12732 	& 17003 (C16)	&	SABm	&	2.18	 	&   	12.3		&	9.56*&	$<$ 38.35					\\
  	ESO 241-G06 	& 17004 (C16)	&	SBm		&	1.54	 	& 	17.3 		&	9.56*&	 $<$ 38.35				\\
	NGC 7793	& 3954 (A)	&	Scd		&	47.75	&	3.9		&	 9.6	 &	$<$ 36.16					\\
	\hline
\end{tabular}
\end{center}
\medskip
Note. -- Columns: (1) Galaxy name; (2) observation ID, denoted as Cycle 16 data (C16) or archival data (A); (3) filtered exposure time in seconds, after correcting for background flares in {\tt CIAO}; (4) morphological type of host galaxy; (5) distance, in Mpc, taken from \cite{Seth2008}; (6) stellar mass of the host galaxy in $M_{\odot}$, taken from \cite{Seth2008}; (7) nuclear X-ray luminosity between 0.3 and 7 keV, in erg s$^{-1}$. \\
${*}$ - denote galaxies with no available V-band data. Galactic mass is taken to be the median value of our late-type sample.
\mbox{}
\end{table*}

\begin{table*}[t]
\begin{center}
\caption{X-ray detections for nucleated late-types}
\label{tab:2}
\begin{tabular}{ccccccccccccc}
	\hline
	\hline 
	\multicolumn{1}{c}{Galaxy} & \multicolumn{1}{c}{Optical $\alpha$ (J200)} & \multicolumn{1}{c}{Optical $\delta$ (J200)} & \multicolumn{1}{c}{$\alpha$ (J200)} & \multicolumn{1}{c}{$\delta$ (J200)} & \multicolumn{1}{c}{$\delta$} & \multicolumn{1}{c}{X-ray counts} \\
	\multicolumn{1}{c}{(1)} & \multicolumn{1}{c}{(2)} & \multicolumn{1}{c}{(3)} & \multicolumn{1}{c}{(4)}  & \multicolumn{1}{c}{(5)} & \multicolumn{1}{c}{(6)} & \multicolumn{1}{c}{(7)} \\
	\hline
	NGC 1042	&	2:40:23.99	&	-8:26:01.1		&	2:40:23.959 (0.12)	&	-8:26:00.57 (0.12)	&	0.6	&	23 (4.8)	\\
	NGC 1493	&	3:57:27.47	&	-46:12:39.2	&	3:57:27.454 (0.10)	&	-46:12:38.55 (0.10)	&	0.6	&	48 (6.9)	\\
	NGC 4487	&	12:31:04.48	&	-8:03:13.8		&	12:31:04.497 (0.12)	&	-08:03:12.48 (0.12)	&	0.8	&	17 (4.1)	\\
	NGC 5879	&	15:09:46.663	&	+57:00:00.67	&	15:09:46.725 (0.08)	&	+57:00:00.64 (0.08)	&	0.2	&	140 (11.8)	\\
	NGC 7690	&	23:33:02.542	&	-51:41:54.11	&	23:33:02.528 (0.13)	&	-51:41:54.44 (0.13)	&	0.3	&	17 (4.1)	\\
	\hline
\end{tabular}
\end{center}
\medskip
Note. -- Units of right ascension are hours, minutes, and seconds, and units of declination are degrees, arcminutes, and arcseconds. Columns: (1) Galaxy name; (2) and (3): R.A. and Dec. of optical center; (4) and (5): R.A. and Dec. of X-ray nucleus with the positional uncertainty on the centroid position given in parenthesis, in arcseconds; (6) $\delta$ between optical center and X-ray nucleus, in arcseconds; (7) nuclear X-ray source net counts extracted between 0.3 and 7 keV, with errors in parenthesis.
\end{table*}

\begin{table*}[t]
\begin{center}
\caption{Nucleated early-type galaxies with X-ray detections}
\label{tab:earlytypes}
\small
\begin{tabular}{ccccccc}
	\hline
	\hline
	\multicolumn{1}{c}{Galaxy} & \multicolumn{1}{c}{Distance} & \multicolumn{1}{c}{log$M_{\ast}$} & \multicolumn{1}{c}{log$L_{\rm X}$}  \\
	\multicolumn{1}{c}{} & \multicolumn{1}{c}{(Mpc)} & \multicolumn{1}{c}{(log$M_{\odot}$)} & \multicolumn{1}{c}{(log(erg s$^{-1}$))} \\
	\multicolumn{1}{c}{(1)} & \multicolumn{1}{c}{(2)} & \multicolumn{1}{c}{(3)} & \multicolumn{1}{c}{(4)} \\
	\hline
	VCC1619	&	15.49	&	10.2 		&	38.68	\\
	VCC1883	&	16.60	&	10.4		&	38.35	\\
	VCC784	&	15.85	&	10.3		&	38.62	\\
	VCC1250	&	17.62	&	10.2		&	38.73	\\
	VCC1283	&	17.38	&	10.1		&	38.54	\\
	VCC1355	&	16.90	&	9.4		&	38.58	\\
	\hline
\end{tabular}
\end{center}

\medskip
Note. -- Columns: (1) Galaxy name; (2) distance, in Mpc; (3) total galactic stellar mass, in log$M_{\odot}$; (4) nuclear X-ray luminosity between 0.3 and 7 keV, in erg s$^{-1}$. All values are taken from \cite{Gallo2010}.

\end{table*}

\begin{table}[t]
\caption{XRB Contamination}
\begin{center}
\label{tab:4}
\small
\begin{tabular}{ccccc}
	\hline
	\hline
	\multicolumn{1}{c}{Galaxy} & \multicolumn{1}{c}{log$L^n_\mathrm{XRB}$} & \multicolumn{1}{c}{$P_\mathrm{SMBH}$} \\
	\multicolumn{1}{c}{} & \multicolumn{1}{c}{(log(erg s$^{-1}$))} & \multicolumn{1}{c}{(\%)}  \\
	\multicolumn{1}{c}{(1)} & \multicolumn{1}{c}{(2)} & \multicolumn{1}{c}{(3)}  \\
	\hline
	N1042	&	37.76	&	94.40		\\
	N1493	&	36.96	&	96.90		\\
	N4487	&	37.10	&	98.07		\\
	N5879	&	37.62	&	96.10		\\
	N7690	&	37.99	&	98.10		\\
	\hline
	\hline
\end{tabular}
\begin{tabular}{ccccc}

	\multicolumn{1}{c}{Galaxy} & \multicolumn{1}{c}{log$L^n_\mathrm{XRB}$} & \multicolumn{1}{c}{$P_\mathrm{SMBH}$} \\
	\multicolumn{1}{c}{} & \multicolumn{1}{c}{(log(erg s$^{-1}$))} & \multicolumn{1}{c}{(\%)}  \\
	\multicolumn{1}{c}{(1)} & \multicolumn{1}{c}{(2)} & \multicolumn{1}{c}{(3)}  \\
	\hline
	V1619	&	37.71	&	92.70		\\
	V1883	&	38.02	&	90.10		\\
	V784		&	38.23	&	92.00		\\
	V1250	&	38.10	&	92.70		\\
	V1283	&	37.73	&	91.80		\\
	V1355	&	36.65	&	92.00		\\	
	\hline
	\hline
\end{tabular}

\medskip
Note. -- Columns: (1) Galaxy name; (2) XRB luminosity between 0.3 to 7 keV expected in central 2$\arcsec$, in erg s$^{-1}$; (3) probability that source is a SMBH instead of an XRB. {\it Top}: late-type sample; {\it Bottom}: early-type sample.
\end{center}
\end{table}

\end{document}